\documentclass[twocolumn,showpacs,prb]{revtex4}

\usepackage{graphicx}
\usepackage{dcolumn}
\usepackage{bm}

\begin{document}

\title{Bose Einstein Condensation in solid $^4$He}
\author{D. E. Galli, M. Rossi and L. Reatto}
\affiliation{
INFM and Dipartimento di Fisica, Universit\`a degli Studi di Milano,
Via Celoria 16, 20133 Milano, Italy}
\date{\today}

\begin{abstract}
We have computed the one--body density matrix $\rho_1$ in solid $^4$He
at $T=0$ K using the Shadow Wave Function (SWF) variational technique.
The accuracy of the SWF has been tested with an exact projector method.
We find that off-diagonal long range order is present
in $\rho_1$ for a perfect hcp and bcc solid $^4$He 
for a range of densities above the melting one, at least up to 54 bars.
This is the first microscopic indication that 
Bose Einstein Condensation (BEC) is present in perfect solid $^4$He.
At melting the condensate fraction in the hcp solid is 
$5 \times 10^{-6}$ and it decreases by increasing the density.
The key process giving rise to BEC is the formation of vacancy--interstitial
pairs.
We also present values for Leggett's upper bound on the superfluid fraction
deduced from the exact local density.
\end{abstract}

\maketitle

The supersolid state, a solid with superfluid properties, is moving out
from theoretical speculations as a result of the observation of non classical
rotational inertia (NCRI) in solid $^4$He in Vycor\cite{chan} 
and very recently in the bulk\cite{cha2}.
The initial theoretical suggestion\cite{andr,ches}
of the supersolid state was based on the possible presence of vacancies
in the ground state of a Bose quantum solid.
In addition these vacancies have to be mobile in order to give rise to Bose Einstein
Condensation (BEC) of $^4$He atoms.
Experimentally no evidence has been found 
for the presence of vacancies at very low temperature
and this is in agreement with the results of microscopic theory\cite{pede,gal5}
which gives an energetic cost of about 15 K for the formation of a vacancy in bulk
solid $^4$He.
However it was almost immediately recognized\cite{legg}
that the presence of ground state vacancies is only one possible mechanism for NCRI
and what is really needed is that atoms are not localized at the lattice sites
but are delocalized via exchange or other processes.
This gives the possibility of having in the wave function (wf)
a phase which governs collectively the motion of the atoms.
The existence of a supersolid state in $^4$He is therefore strictly
related to the question of localization or delocalization of particles
and, of course, this is a topic of general interest.
This is the case, for instance, of
cold alkali atoms in a periodic potential\cite{grei}.
Experiments with $^4$He give access to the superfluid fraction $\rho_s/\rho$ which turns out
to be\cite{cha2} at most of order 2\%.
Theoretically only upper bounds on $\rho_s$ have been obtained up to now\cite{puma,sasl,sas2}
and no microscopic theory has given evidence for a supersolid phase.
The commonly accepted view is that one can have a finite $\rho_s$
in a three dimensional system if there is BEC in the system so a central
quantity to compute is the off diagonal one--body density matrix
$\rho_1({\vec r},{\vec r}\,')$, whose
Fourier transform represents the momentum distribution.

In this article we address the computation of $\rho_1$
for solid $^4$He at $T=0$ K based on a variational wf,
a shadow wf (SWF). In a previous computation\cite{gal1} we have found
that the presence of vacancies in the solid
induces a BEC which is proportional to the concentration of vacancies.
On the other hand the large energy of formation of a vacancy
makes the probability of having such defects at low
temperature vanishing small.
Here we study in the perfect solid the large distance behavior of
$\rho_1({\vec r},{\vec r}\,')$, specifically if
$\rho_1$ has a non zero limit at large distance (off diagonal
long range order, ODLRO) which implies BEC.
With perfect solid we mean that the number of maxima in the local density $\rho({\vec r})$
is equal to the number of $^4$He atoms.
The use of SWF is specially useful in the present context because with
such wf the crystalline order is an effect of the spontaneously
broken symmetry so that local disorder processes like exchange of two
or more particles, creation of vacancy--interstitial pairs (VIP) or more
complex processes are in principle allowed.
The major finding of our computation is the presence of a small but
finite condensate for a range of densities above melting.
The variational theory is very useful to describe strongly 
interacting systems like liquid or
solid $^4$He but it is always open to debate how much the results
depend on the ansatz on the wf, specially for quantities other than the energy.
In order to give indication on the reliability of our SWF we present
some results on quantities like the degree of local order,
of localization and of the local density obtained also
from an exact computation based\cite{gal3} on a projection
algorithm (SPIGS), a Path Integral Ground State\cite{pigs} method
which uses a SWF as the starting wf.

In a SWF the correlations between atoms are introduced both explicitly 
by a Jastrow factor and also in an implicit
way by coupling with a set of subsidiary variables, which are called ``shadow''
variables\cite{viti} (one shadow for each quantum particle),
which are integrated over.
All expectation values
are computed by a Monte Carlo (MC) method 
and the statistical sampling of $|\Psi|^2$
maps the quantum system of $N$ particles in a system of $N$ special interacting
triatomic ``molecules''\cite{viti} which consist of a $^4$He atom and two shadows.
The accuracy of the SWF technique is well documented and
it has been possible to treat also disorder phenomena in a quantum solid,
like a vacancy\cite{pede,gal5} 
or even the interfacial region between a solid and a
liquid at coexistence\cite{ferr}.
As functional form for the correlating factors contained in the SWF
we have taken the ones used
in ref.\cite{pede}; as interatomic interaction we have used
a standard Aziz potential\cite{azi1}.

A SWF can be interpreted as a first projection step in imaginary time
of a Jastrow wf via a variationally
optimized imaginary time propagator\cite{gal3}.
With SPIGS one
goes beyond the variational theory by
adding successive projection steps
in the imaginary time propagation with the full Hamiltonian
and in this way we are able to
compute exact expectation values on the ground state without extrapolations.
With these two Quantum MC methods no a priori equilibrium positions for the solid phase are
required, the Bose symmetry is manifestly maintained and atoms 
can be delocalized. 

The equation of state given by SWF is in good agreement with the results of SPIGS,
for instance at melting ($\rho=0.029$ \AA$^{-3}$) the energy per particle is -5.12 K,
0.76 K above the SPIGS result\cite{moro}. Also for a vacancy there is an excellent agreement,
the formation energy at melting is 15.7 K with SPIGS and 15.6 K with SWF\cite{gal5} and the jumping 
rate is similar in the two computations.
In addition to the energy one would like to know the accuracy of SWF in describing
the microscopic local processes and to this end we have computed the local density
$\rho({\vec r})$ and the static structure factor $S({\vec k})$.
We find again agreement between SWF and SPIGS with SWF giving a slightly more ordered state.
For instance at melting 
the main Bragg peak of the hcp solid is about 17\% higher than the SPIGS result.
Detailed comparison will be presented elsewhere and here we focus only on the results for 
the upper bound $f_s^+$ for the superfluid fraction $\rho_s/\rho$ obtained by Leggett\cite{legg}.
This bound depends on
the averaged density, $\rho(z)=\int d\xi \rho({\vec r})$, where
$z$ is a longitudinal coordinate and
$\xi$ is a suitable set of transversal coordinates\cite{legg}.
We have chosen as $z$ axis the one which gives the lowest value of $f_s^+$;
in the hcp crystal
this is the $\Gamma$A direction which is perpendicular to the basal plane.
It should be noticed that usually in Quantum MC computations the center of
mass of the system is not fixed and this would alter the local density specially
around the minima. Therefore when we compute $\rho({\vec r})$ we have modified the
sampling algorithm to keep
the center of mass of the system fixed.
\begin{table}
 \caption{\label{tab} Upper bounds $f_s^+$ for the superfluid fraction in hcp bulk solid $^4$He
          computed at different densities with the SWF
          technique, the SPIGS method and the GM for $\rho({\vec r})$.
          $P$ is the pressure from the SPIGS equation of state.
          $\sigma$ is the standard deviation of the Gaussians used in the GM.}
 \begin{ruledtabular}
  \begin{tabular}{cccccccc}
 & & & SWF & SPIGS & & GM & \\
 $\rho$ (\AA$^{-3}$) & $P$ (bar) & &
                      $f_s^+$&$f_s^+$& &$f_s^+$&$\sigma$ (\AA) \\ \hline
    0.0290 &  29.3 & & 0.287 & 0.384 & & 0.380 & 0.543  \\
    0.0310 &  53.6 & & 0.255 & 0.299 & & 0.297 & 0.503  \\
    0.0330 &  87.8 & & 0.209 & 0.230 & & 0.222 & 0.467  \\
    0.0353 & 141.9 & & 0.141 & 0.164 & & 0.166 & 0.436  \\
    0.0400 & 316.9 & & 0.077 & 0.080 & & 0.079 & 0.381  \\
    0.0440 & 553.5 & & 0.042 & 0.042 & & 0.041 & 0.345  \\
  \end{tabular}
 \end{ruledtabular}
\end{table}
In Tab.\ref{tab} we show the upper bounds, $f_s^+$, for the superfluid fraction
obtained for hcp bulk solid $^4$He with the SPIGS and SWF methods.
There is substantial agreement, the variational $f_s^+$
being always lower as a consequence of the larger degree of local order.
A popular representation of $\rho({\vec r})$ is the one in terms of sum of Gaussians
centered on the lattice sites. We have fitted our $\rho({\vec r})$ with this Gaussian
Model (GM) by using the standard deviation $\sigma$ as fitting parameter.
We find that the GM gives an excellent representation of the integrated density along
planes, what is needed in the Leggett's inequality, the deviation being below
4\%. The resulting bounds given by GM with $\sigma$ fitted on the SPIGS $\rho({\vec r})$
are also shown in Tab.\ref{tab}.
The bound $f_s^+$ given by SWF is similar to the value computed previously\cite{puma}
with the GM fitted on a different variational theory.
In Ref.\cite{sasl,sas2} a lower $f_s^+$ has been obtained by using a better variational ansatz:
the phase of the wf is a function of ${\vec r}$ and not only of the
longitudinal coordinate $z$ as in Ref.\cite{legg}.
This bound computed
with the GM for a given crystal lattice is simply
a function of the ``localization parameter''\cite{sasl,sas2}.
Saslow's computation is for an fcc crystal but if 
we neglect the difference between the hcp and the fcc lattice,
using the $\sigma$ in Tab.\ref{tab} and the results in Ref.\cite{sas2}, 
we can estimate an $f_s^+$ which goes from about
0.2 at $\rho=0.029$\AA$^{-3}$
to about 0.005 at $\rho=0.044$\AA$^{-3}$ .
A SPIGS computation for fcc at $\rho=0.029$\AA$^{-3}$
gives a $\sigma$ in the GM which is only 2\% lower than in hcp crystal.
These values of $f_s^+$ are compatible with the experiments but $f_s^+$  is about one order of
magnitude larger than the experimental value of $\rho_s/\rho$ and, in any case,
it is only an upper bound so it is not very conclusive.
A word of caution on the GM is in order. If this model gives an excellent representation
for the integrated density $\rho(z)$, the accuracy is lost when we consider the local
density $\rho({\vec r})$: in the region of the minima of $\rho({\vec r})$ deviations
greater than 100\% are found.

The one--body density matrix
$\rho_1({\vec r},{\vec r}\,')$
is given by the
overlap between the normalized many--body
ground state wf $\Psi(R)$ and $\Psi(R')$ where
configuration $R'=\{ {\vec r}\,',{\vec r}_2,..,{\vec r}_N\}$ differs from 
$R =\{ {\vec r},{\vec r}_2,..,{\vec r}_N\}$ only by the position of one 
of the $N$ atoms in the system;
if $\Psi(R)$ is translationally invariant as in our case when the center of mass is not fixed, 
$\rho_1$ only depends on the difference ${\vec r}-{\vec r}\,'$:
\begin{equation}
\rho_1({\vec r}-{\vec r}\,')= N \int d{\vec r}_2 .. d{\vec r}_N
\Psi^*(R)
\Psi(R')\quad . \label{due}
\end{equation}
It is possible to interpret the integrand in Eq.(\ref{due}) as a probability density\cite{gal1};
then $\rho_1$ can be computed by sampling the integrand in
Eq.(\ref{due}) and by histogramming the occurrence of the distance ${\vec d}=\vec{r}-\vec{r}\,'$.
In the following we will call ``half'' particles
the particles with coordinates ${\vec r}$ and ${\vec r}\,'$
because they have just ${1 \over 2}$ the correlation strength
with the other $N-1$ particles (with coordinates $\{{\vec r}_2,..,{\vec r}_N\}$)
and no direct correlation between them.
The method of computation has been described in Ref.\cite{gal1}.
Absence of ODLRO corresponds to the two ``half'' particles forming a ``molecule''
whereas presence of ODLRO corresponds to a finite probability of dissociation
up to infinite distances.

\begin{figure}
 \includegraphics[angle=0, width=8.5 cm]{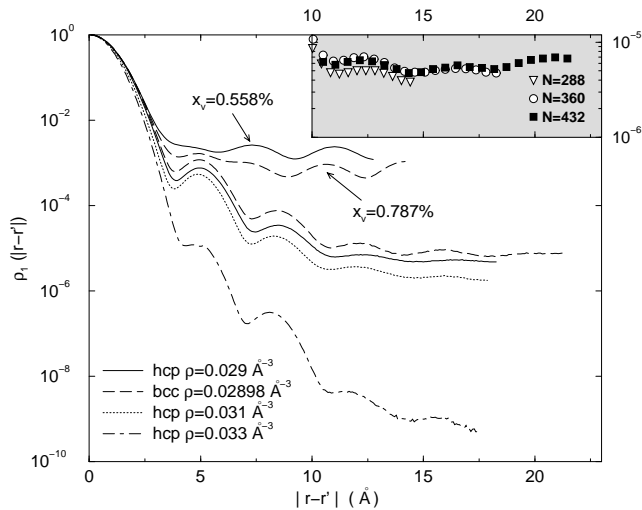}
  \caption{\label{fig:odlro} $\rho_1(\vec{r}-\vec{r}\,')$ at
           different densities for hcp and bcc perfect $^4$He crystals and for
           the same crystals with a finite concentration $x_{\rm v}$ of vacancies.}
\end{figure}

We have computed
$\rho_1(\vec{r}-\vec{r}\,')$ along the nearest neighbors (nn) direction
which is $\Gamma$K in a hcp  and [111] in a bcc crystal.
In Fig.\ref{fig:odlro} we report $\rho_1$ for
a perfect hcp crystal at different densities at and above melting and 
for a bcc crystal at $\rho =0.02898$\AA$^{-3}$. 
It is clear that at melting and at $\rho =0.031$\AA$^{-3}$
$\rho_1$ reaches a plateau at large distance whereas at $\rho =0.033$\AA$^{-3}$
$\rho_1$ steadily decreases with increasing distance.
By averaging the tail in $\rho_1$ for distances greater than 14 \AA,
we find at the melting density a condensate fraction
$n_c=(5.0 \pm 1.7)\times 10^{-6}$ in a perfect hcp
and $(7.6 \pm 1.7)\times 10^{-6}$ in a perfect bcc crystal.
This is the first microscopic indication that 
BEC is present in perfect solid $^4$He\cite{not2}.
At $\rho =0.031$\AA$^{-3}$ we find $n_c=(2.0 \pm 0.4) \times 10^{-6}$ and
at $\rho =0.033$\AA$^{-3}$ the tail is so much depressed that
the size of the simulation box is too small to conclude if ODLRO is present;
in this case we can only say that the condensate fraction, if any, is lower
than $10^{-9}$.
The simulation box which has been used to compute $\rho_1$ is cubic and contains $N=432$ $^4$He
atoms for the bcc crystal, it is elonged in the $\Gamma$K direction and contains 
$N=360$ $^4$He atoms for the hcp crystal.
By changing the size of the boxes at fixed density we have checked that our results for 
$\rho_1$ have no finte size effect within the statistical errors of our computations.
This is shown in the inset of Fig.\ref{fig:odlro} for the hcp crystal at $\rho =0.029$\AA$^{-3}$;
by averaging the tail in $\rho_1$ for distances greater than 14 \AA,
we find $n_c=(3.9 \pm 1.7)\times 10^{-6}$ for $N=288$ and $n_c=(5.7 \pm 2.0)\times 10^{-6}$
for $N=432$, to be compared with the value given above: $n_c=(5.0 \pm 1.7)\times 10^{-6}$
for $N=360$.
In ref.\cite{gal1} we found that
a finite concentration of vacancies $x_{\rm v}$ induces a condensate fraction
which depends linearly on $x_{\rm v}$.
In Fig.\ref{fig:odlro} we show also $\rho_1$ when a vacancy is present both in hcp\cite{gal1}
and bcc.
Taking into account the value of $x_{\rm v}$ of the computation we estimate
that at $\rho =0.029$\AA$^{-3}$ the condensate fraction due to a
finite concentration of vacancies
is equal to the one in a perfect hcp crystal when $x_{\rm v}\simeq 1.5\times 10^{-5}$.

\begin{figure}
 \includegraphics[angle=0, width=8 cm]{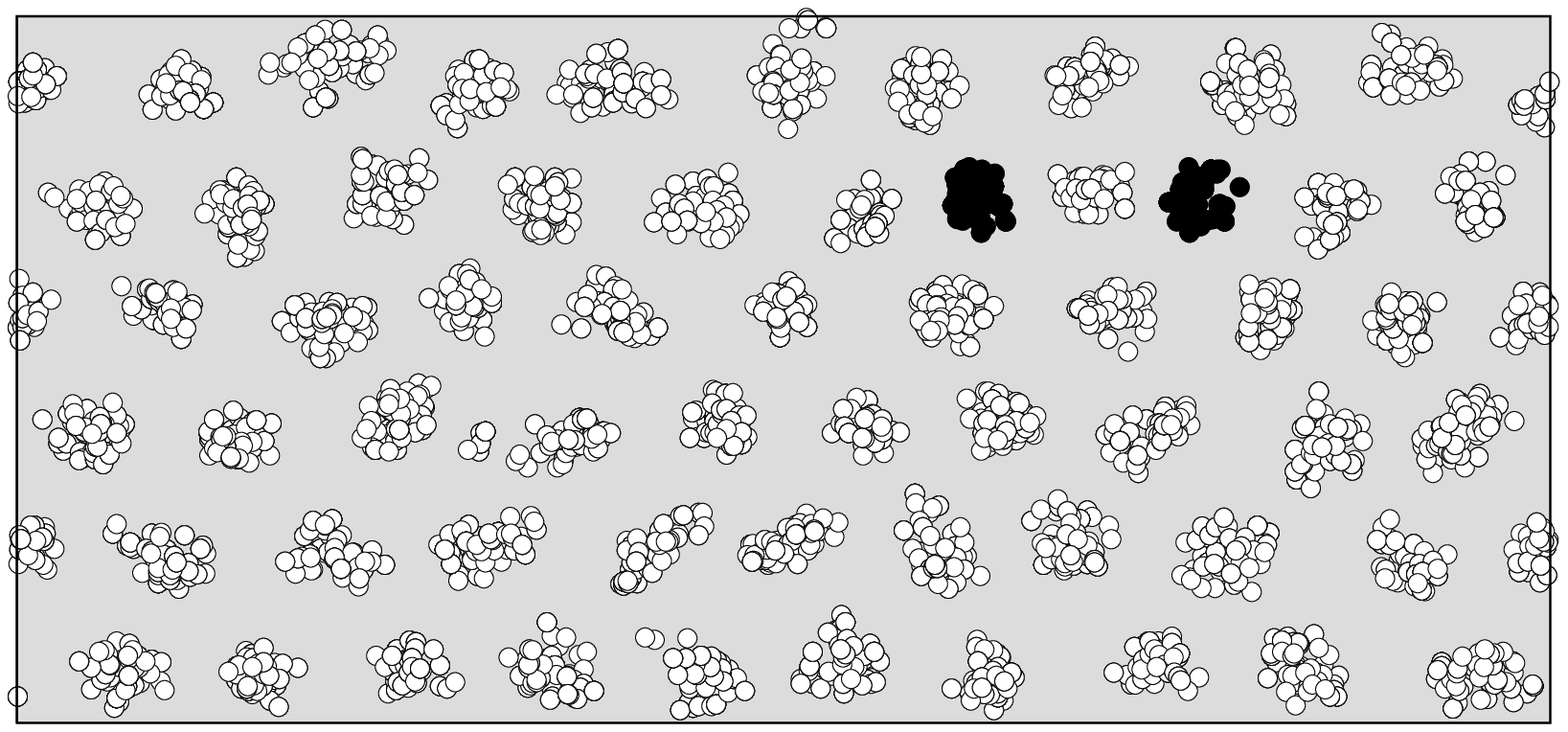}
  \caption{\label{configu} Projection of 100 successive configurations
          of the particles (white circles)
          and the two ``half'' particles (black circles)
          in a basal plane of an hcp crystal at $\rho =0.031$\AA$^{-3}$.
          $\rho_1$ is the probability distribution of the two ``half'' particles.
           }
\end{figure}

All the computed $\rho_1$ show oscillations which reflect the crystalline order in the system.
However these oscillations are not the same in the perfect and in the defected solid.
We find that when a vacancy is present
the maxima of the oscillations in $\rho_1$ correspond to
multiples of the nn distance $d_{\rm nn}$.
This is an indication that in presence of a vacancy the main mechanism which
contributes to the separation of the two ``half'' particles is that
one of them moves through
the crystal following the vacancy which is very mobile\cite{gal2}.
The different positions of the maxima in $\rho_1$ for the perfect crystal 
suggest a different microscopic process for the ODLRO in this case.
By analysing the particle configurations sampled in our runs we find that the secondary
peak of $\rho_1$, located at about 5 \AA , always corresponds to a configuration
in which the two ``half'' particles occupy two nn lattice positions slightly
distorted by the presence of one interstitial $^4$He atom between them.
In Fig.\ref{configu} we show 100 successive configurations of the particles which corresponds
to this event.
In the formalism of second quantization $\rho_1(\vec{r}-\vec{r}\,')$ is
equal to the
expectation value of the composite event where one $^4$He atom is destroyed at ${\vec r}\,'$
and one is created at ${\vec r}$; then it is possible to interpret the event in Fig.\ref{configu}
as the creation of a VIP.
The same process is found in bcc crystal. After this first step the two ``half''
particles have a finite probability of moving away one from another by exchange
processes with the other atoms and this gives rise to the other maxima of $\rho_1$
at larger distance. By analysing the particle configurations corresponding to these other maxima
we find that a VIP is present in all the configurations.
Similar processes were considered in Ref.\cite{pro} as a necessary condition
for the supersolid phase, but there it is argued that VIP cannot be present.
Our results disagree with this hypothesis.
In order to characterize the anisotropy of $\rho_1$ as function of ${\vec d}$
we have computed $\rho_1$ 
when the two ``half'' particles are no more constrained to lie on the nn direction
but can freely move in a plane.
\begin{figure}
 \includegraphics[angle=0, width=8 cm]{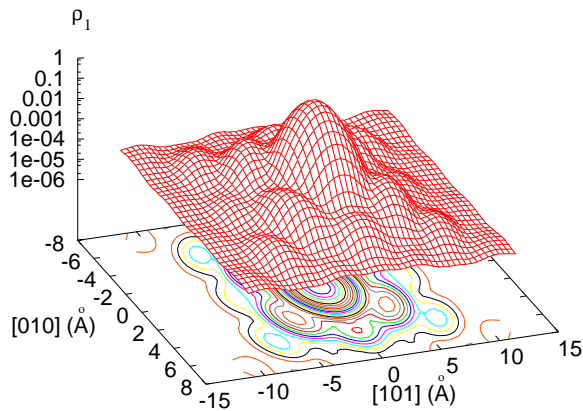}
  \caption{\label{piano}
           $\rho_1(\vec{r}-\vec{r}\,')$ at
           $\rho=0.02898$\AA$^{-3}$ for a perfect bcc crystal with ${\vec r}-{\vec r}\,'$
           lying in the plane [-101].
           }
\end{figure}
In Fig.\ref{piano} one can see that $\rho_1$ in a perfect bcc crystal is strongly anisotropic
for distances up to about 6 \AA, and the maxima of $\rho_1$
are in the direction of nn.
However at greater distances $\rho_1$ becomes nearly isotropic and we
conclude that our estimation of the BEC fraction 
is not affected by the previous restriction on ${\vec d}$.
Similar results are obtained for hcp.
It is interesting to notice that
when a vacancy is present the anisotropy of $\rho_1$
persists up to greater distances (data not shown).
Also in Fig.\ref{fig:odlro} one can see that
the oscillations of $\rho_1$ are more persistent with increasing distance in the
crystal with a vacancy;
this is another indication that
different microscopic processes are at the origin
of the ODLRO in the perfect and in the
defected solid $^4$He.
The exchange of atoms and VIPs are present not only in
$\rho_1$ but also in $|\Psi|^2$. At melting about every $2\times 10^3$ MC steps
an atom has a displacement larger than $d_{\rm nn}$ and in many cases this is associated
with the presence of an interstitial.
In principle one can devise an algorithm based on SPIGS to compute $\rho_1$ exactly.
However at present this appears to be a major computational problem.
In any case we have given solid evidence that SWF overestimates the degree of local order
so that we should expect that the SWF results for the BEC fraction are an underestimation of
the exact values.

In conclusion we have shown that solid $^4$He at $T=0$ K has BEC at 
melting density and
above at least up to 54 bars whereas we find a vanishing BEC at 90 bars.
Thus BEC should be at the basis of the NCRI observed experimentally\cite{cha2}.
Our result has been obtained from an advanced variational theory the accuracy
of which has been tested with a projector method on the exact ground state.
The key process giving rise to ODLRO is the formation of a VIP.
Such defects have a finite probability to be present in the ground state of the system;
they are not permanent excitations
but simply rare fluctuations of the perfect crystal induced by
the large zero--point motion.
In other words the number of atoms is equal to the number of lattice sites
and, at the same time, atoms are delocalized.
Since the ground state is the vacuum of the elementary excitations of the system we
conjecture that a branch of low energy excitations different from phonons should be present
in solid $^4$He.
Such excitations should have an important role in determining the critical temperature.
It is a possibility that this branch is related to some experimental results which have
been interpreted in term of an excitation with energy of about 2 K\cite{good}.

This work was supported by the INFM Parallel Computing
Initiative and by the Mathematics Department ``F. Enriques''
of the Universit\`a degli Studi di Milano.


\begin{thebibliography}{99}
\bibitem{chan} E. Kim, M.H.W. Chan, {\it Nature} {\bf 427}, 225 (2004).
\bibitem{cha2} E. Kim, M.H.W. Chan, {\it Science} {\bf 305}, 1941 (2004).
\bibitem{andr} A.F. Andreev and I.M. Lifshitz, {\it Soviet Phys. JETP}
{\bf 29}, 1107 (1969).
\bibitem{ches} G.V. Chester, {\it Phys. Rev. A} {\bf 2}, 256 (1970).
\bibitem{pede} F. Pederiva, G.V. Chester, S. Fantoni and L. Reatto
{\it Phys. Rev. B} {\bf 56} {5909}, (1997).
\bibitem{gal5} D.E. Galli, L. Reatto, {\it J. Low Temp. Phys.} {\bf 134}, 121 (2004).
\bibitem{legg} A.J. Leggett, {\it Phys. Rev. Lett.} {\bf 25}, 1543 (1970);
A.J. Leggett, {\it J. Stat. Phys.} {\bf 93}, 927 (1998).
\bibitem{grei} M. Greiner, O. Mandel, T. Esslinger, T.W. Hansch, I. Bloch, {\it Nature} {\bf 415}, 39 (2002).
\bibitem{puma} J.F. Fernandez, M Puma, {\it J. Low Temp. Phys.} {\bf 17}, 131 (1974).
\bibitem{sasl} W.M. Saslow, {\it Phys. Rev. Lett.} {\bf 36}, 1151 (1975).
\bibitem{sas2} W.M. Saslow, cond-mat/0407166.
\bibitem{gal1} D.E. Galli, L. Reatto, {\it J. Low Temp. Phys.} {\bf 124}, 197 (2001).
\bibitem{gal3} D.E. Galli, L. Reatto, {\it Mol. Phys.} {\bf 101}, 1697 (2003).
\bibitem{pigs} A. Sarsa, K.E. Schmidt, W.R. Magro, {\it J. Chem. Phys.} {\bf 113}, 1366 (2000).
\bibitem{viti} S.A. Vitiello, K. Runge, M.H. Kalos, {\it Phys. Rev. Lett.} {\bf 60}, 1970 (1988).
\bibitem{ferr} F. Pederiva, A. Ferrante, S. Fantoni and L. Reatto
{\it Phys. Rev. Lett.} {\bf 72} {2589}, (1994).
\bibitem{azi1} R.A. Aziz, V.P.S. Nain, J.S. Carley, W.L. Taylor, G.T. McConville,
{\it J. Chem. Phys.} {\bf 70}, 4330 (1979); 
some tests with more recent forms for the interatomic potential
give results similar to the present ones.
\bibitem {moro} This difference is reduced to 0.31 K with a fully optimized SWF
(S. Moroni, D.E. Galli, S. Fantoni and L. Reatto, {\it Phys. Rev. B} {\bf 58}, 909 (1998))
but that optimization was performed only over a restricted density range
so we have used the older form of SWF.
\bibitem{not2} In Ref.\cite{gal1} we concluded that there was no BEC in the perfect solid,
this was due to equilibration problems.
\bibitem{gal2} D.E. Galli, L. Reatto, {\it Phys. Rev. Lett.} {\bf 90}, 175301 (2003).
\bibitem{pro} N. Prokof'ev and B. Svistunov, cond-mat/0409472.
\bibitem{good} J.M. Goodkind,  {\it Phys. Rev. Lett.} {\bf 89}, 095301 (2002).
\end{thebibliography}
\end{document}